\documentclass[aps,prl,floatfix,showpacs,twocolumn]{revtex4}
\usepackage{amsmath,amssymb}
\usepackage{graphicx}
\usepackage{epsfig}

\newcommand{\bm}{\boldsymbol}

\begin{document}

\hsize\textwidth\columnwidth\hsize\csname@twocolumnfalse\endcsname

\title{Energy Relaxation of Hot Dirac Fermions in Graphene}

\author{Wang-Kong Tse$^{1,2}$}
\author{S. Das Sarma$^1$}
\affiliation{$^1$Condensed Matter Theory Center, Department of Physics,
University of Maryland, College Park, Maryland 20742, USA}
\affiliation{$^2$Department of Physics, University of Texas, Austin, Texas 78712, USA}

\begin{abstract}
We develop a theory for the energy relaxation of hot Dirac fermions in graphene. We obtain a generic expression for the energy relaxation rate due to electron-phonon interaction and calculate the power loss due to both optical and acoustic phonon emission as a function of electron temperature $T_{\mathrm{e}}$ and density $n$. We find an intrinsic power loss weakly dependent on carrier density and non-vanishing at the Dirac point $n = 0$, originating from interband electron-optical phonon scattering by the intrinsic electrons in the graphene valence band. We obtain the total power loss per carrier $\sim 10^{-12}\,-\,10^{-7}\,\mathrm{W}$ within the range of electron temperatures $\sim 20\,-\,1000\,\mathrm{K}$. We find optical (acoustic) phonon emission to dominate the energy loss for $T_{\mathrm{e}} > (<)\,200-300\,\mathrm{K}$ in the density range $n = 10^{11}-10^{13}\,\mathrm{cm}^{-2}$. 
%
\end{abstract}


\maketitle

Graphene is a two-dimensional (2D) plane of carbon atoms in a hexagonal lattice, with the motion of the electrons near the Brillouin zone K point (the `Dirac point') obeying the massless Dirac equation. This unusual quasi-relativistic property implies a linear energy dispersion $\epsilon_{k\lambda} = \hbar\lambda vk$ of the conduction band ($\lambda = 1$) and the valence band ($\lambda = -1$) intersecting at the Dirac point, with a constant Fermi velocity $v = 10^6\,\mathrm{ms}^{-1}$. The recent experimental discovery of this new 2D material has sparked enormous interest in understanding its fundamental transport properties; for low electric fields, carrier transport is determined by the behavior of the carriers close to the Fermi level, and has been quite extensively studied both experimentally and theoretically \cite{GrapheneTransport}. 

What still remains as an interesting open arena for exploration is the carrier transport in graphene under high electric fields and in non-equilibrium (i.e. ``hot-electron'') conditions. With the application of a high electric field, carriers gain energy at a rate much faster than that for the carriers to lose energy to the lattice, creating a non-equilbrium carrier population which subsequently comes to an internal thermal equilibrium (among the carriers themselves but not with the lattice) through carrier-carrier scattering. These carriers (called hot carriers), having a higher temperature $T_{\mathrm{e}}$ than the lattice temperature $T_{\mathrm{L}}$, will then relax towards a thermal equilibrium with the lattice by losing energy to the lattice. As transport properties at high fields are determined by these hot carriers, a quantitative understanding of this hot carrier energy relaxation process is therefore of central importance in high-field carrier transport; furthermore, it is a crucial issue affecting the performance characteristics of ultrafast, high-field devices. 

Experimentally, this energy relaxation process can be studied, following excitation with an ultrashort laser pulse, by measuring the transmission spectrum \cite{Shank,DeHeer,Spencer} (pump-probe spectroscopy) or the luminescence spectrum \cite{Kash} (photoluminescence spectroscopy) from the sample. In particular, the luminescence spectrum yields directly the temperature of the hot carriers, and when combined with electrical (instead of optical) excitation of the carriers with a steady-state electric current, can provide a direct means to measure the carrier temperature as a function of the energy-loss rate \cite{Shah_DC}. 

In this Letter, we develop a theory for the energy relaxation of hot electrons in graphene, incorporating the energy loss channels due to optical phonon emission and acoustic phonon emission. Note that electron-electron interaction can only lead to energy equilibration within the carrier system (i.e. establishing an electron temperature), but cannot contribute to the overall energy loss from the carriers to the lattice. We derive an expression for the energy relaxation rate (i.e., power loss) of hot electrons in graphene due to electron-phonon (e-ph) interaction, and obtain the power loss as a function of electron temperature and density. We find an intrinsic power loss that is present at all doping levels and does not vanish at zero doping arising entirely due to interband electron-optical phonon scattering by the intrinsic electrons in the valence band; and an extrinsic power loss which scales with doping density. We also find that the power loss is predominantly due to acoustic phonon emission below $\sim 200\,-\,300\,\mathrm{K}$ (depending on the doping density) with optical phonon emission taking over as the dominant energy loss mechanism above this temperature range. The temperature throughout this paper refers to the electron temperature with the lattice assumed to be held at a low temperature.

The rates of change of the electron and hole distribution functions describing the electron-phonon scattering are given by ($\bm{r}$ and $t$ dependence of the distribution functions are suppressed for clarity, and we set $\hbar = 1$ throughout unless specified):
\begin{widetext}
\begin{eqnarray}
&&\left(\frac{\partial f_{k\mu}}{\partial t}\right)_{\mathrm{col}} = -2\pi\sum_{q,\lambda}\alpha_{{k},{k}-{q}}^{\mu\lambda}\left\{\left[\left(N_q+1\right)\delta\left(\xi_{k\mu}-\xi_{k-q\lambda}-\omega_q\right)+N_q\delta\left(\xi_{k\mu}-\xi_{k-q\lambda}+\omega_q\right)\right]f_{k\mu}\left(1-f_{k-q\lambda}\right)\right. \nonumber \\
&&\left.-\left[\left(N_q+1\right)\delta\left(\xi_{k\mu}-\xi_{k-q\lambda}+\omega_q\right)+N_q\delta\left(\xi_{k\mu}-\xi_{k-q\lambda}-\omega_q\right)\right]\left(1-f_{k\mu}\right)f_{k-q\lambda}\right\}. \label{eq1}
\end{eqnarray}
\end{widetext}
where $\mu,\lambda = \pm 1$ is the chirality index for electron ($\mu,\lambda = 1$) in the conduction band or hole ($\mu,\lambda = -1$) in the valence band, $f_{k\mu}$ is the electron ($\mu = 1$) or hole ($\mu = -1$) distribution function \cite{Remark1}, $\xi_{k\lambda} = \lambda vk-\mu_{\mathrm{c}}$ is the quasiparticle energy rendered from the chemical potential $\mu_{\mathrm{c}}$, $N_q$ is the phonon distribution function at the phonon energy $\omega_q$, $\alpha_{k,k-q}^{\mu\lambda}$ is the e-ph coupling strength which, for graphene, has a non-trivial chiral and momentum dependence arising from the graphene band structure and the e-ph interaction vertex \cite{Ando,myphonon1,myphonon2}. In Eq.~(\ref{eq1}), the first term within the brackets on the right hand side describes the scattering of an electron in the state $(\bm{k},\mu)$ into another state $(\bm{k}-\bm{q},\lambda)$ via the emission (i.e., $\xi_{k\mu}-\xi_{k-q\lambda} = \omega_q$) or absorption ($\xi_{k-q\lambda}-\xi_{k\mu} = \omega_q$) of a phonon, and the second term follows with an analogous physical meaning. The sum over $\lambda = \pm 1$ takes into account both intraband ($\lambda = \mu$) and interband ($\lambda \ne \mu$) scattering processes through phonon emission or absorption. 

We now state the assumptions for our model: (1) After the initial rapid carrier-carrier scattering, the electron gas has established an internal thermal equilibrium at an electron temperature $T_{\mathrm{e}}$ described by the Fermi distribution function $f_{k\mu} = n_F(\xi_{k\mu})$, where $n_F(\varepsilon) = 1/[\mathrm{exp}(\varepsilon/k_{\mathrm{B}}T_{\mathrm{e}})+1]$ is the Fermi function. (2) We take into account the fact that the emitted optical phonons can decay into low-energy acoustic phonons due to anharmonic phonon-phonon scattering, which is characterized phenomenologically by a finite optical phonon lifetime $\tau_{\mathrm{ah}}$. 
(3) Acoustic phonons emitted by the electrons or produced through the decay of optical phonons thermalize immediately with the lattice (acting as a heat bath) which is maintained at a lattice temperature $T_{\mathrm{L}} < T_{\mathrm{e}}$. Recent ultrafast optical spectroscopy experiments on graphene \cite{DeHeer,Spencer} find that the time for the electrons to equilibrate among themselves spans $\sim 100\,\mathrm{fs}$, and the subsequent thermalization of the electron gas with the lattice lasts for $\sim 1\,\mathrm{ps}$. It is within this picosecond time scale that the electrons lose most of their energy through e-ph scattering as electron-electron scattering does not dissipate energy from the electron gas as a whole. 

We first ignore the effect of a finite phonon lifetime by taking $\tau_{\mathrm{ah}} = 0$, assuming that the emitted phonons immediately thermalize with the lattice and all the energy lost from the electrons to the phonons is also immediately lost to the lattice. 
The energy loss rate $\mathrm{d}E_{k\mu}/\mathrm{d}t$ of a single electron with momentum $\bm{k}$ can be obtained from Eq.~(\ref{eq1}) by inserting the energy change $\xi_{k-q\lambda}-\xi_{k\mu}$ of the electron due to scattering under the $\bm{q}$ integral. The total energy loss rate $P$ of the entire system of electrons then follows by summing the resulting single-electron energy loss rate over all states of momentum $\bm{k}$ and chirality $\mu$, taking into account the degeneracy factors due to spins $g_{\mathrm{s}} = 2$ and valleys $g_{\mathrm{v}} = 2$ to give $P = -g_{\mathrm{v}}g_{\mathrm{s}}\sum_{\mu}\sum_{{k}}{\mathrm{d}E_{k\mu}}/{\mathrm{d}t}$.
%
%
Making use of the integral identities $\delta(\xi_{k\mu}-\xi_{k-q\lambda}\mp\omega_q) = \int\mathrm{d}\omega\delta(\xi_{k\mu}-\xi_{k-q\lambda}+\omega)\delta(\omega_q\pm\omega)$ in the expression of ${\mathrm{d}E_{k\mu}}/{\mathrm{d}t}$ obtained above, 
we arrive, after some algebra, at the following expression for the total power loss from the electrons: 
\begin{eqnarray}
P &=& 2\sum_{{q}}\int_{-\infty}^{\infty}\frac{\mathrm{d}\omega}{\pi}\omega\left[n_{B}^{\mathrm{L}}(\omega)-n_{B}^{\mathrm{e}}(\omega)\right] \nonumber \\
&&\times\mathrm{Im}\Pi^{\mathrm{ph}}({q},\omega)\mathrm{Im}\mathcal{D}({q},\omega),
\label{eq2}
\end{eqnarray}
where $n_{B}^{\mathrm{e,L}}(\omega) =  1/[\mathrm{exp}(\omega/k_{\mathrm{B}}T_{\mathrm{e,L}})-1]$ stand for the Bose distribution functions evaluated at the electron $T_{\mathrm{e}}$ and lattice $T_{\mathrm{L}}$ temperatures respectively, $\mathcal{D}(\bm{q},\omega) = 2\omega_q/(\omega^2-\omega_q^2+i0^+)$ is the phonon Green function, 
\begin{eqnarray}
\Pi^{\mathrm{ph}}(q,\omega) = g_{\mathrm{v}}g_{\mathrm{s}}\sum_{q\mu\lambda}\alpha_{{k},{k}-{q}}^{\mu\lambda}\frac{n_F(\xi_{k\mu})-n_F(\xi_{k-q\lambda})}{\omega+\xi_{k\mu}-\xi_{k-q\lambda}+i0^+}, 
\label{eq3}
\end{eqnarray}
is the phonon self-energy \cite{myphonon2} at the electron temperature $T_{\mathrm{e}}$, and `$\mathrm{Im}$' in Eq.~(\ref{eq2}) stands for the imaginary part.  
Eq.~(\ref{eq2}) generalizes the Kogan formula \cite{Kogan} for the power loss in an e-ph coupled system widely used in regular metals and semiconductors \cite{2D_Th} with a parabolic energy band to a chiral two-band system (to which graphene belongs as a special case), embodying both intraband and interband electronic transitions as well as the non-trivial chiral and momentum dependence of the e-ph coupling.  
%

We first consider the energy relaxation due to optical phonon emission. The LO phonon mode at the Brillouin zone center $\Gamma$ in graphene is characterized by the phonon energy $\omega_0 = 196\,\mathrm{meV}$ and e-ph coupling $\alpha_{{k},{k}-{q}}^{\mu\lambda} = g_{\mathrm{op}}^2[1-\mu\lambda\mathrm{cos}(\phi_k+\phi_{k-q}-2\phi_q)]/2$ \cite{Ando,myphonon1,myphonon2}, where $g_{\mathrm{op}}^2$ is the e-ph coupling constant \cite{Remark3} and $\phi_k = \mathrm{tan}^{-1}(k_y/k_x)$. The imaginary part of the phonon self-energy Eq.~(\ref{eq3}) describes the damping of the phonon mode due to electron-hole pair excitations. For the $\Gamma$ point optical phonons in graphene, $\Pi^{\mathrm{ph}}$ is different from the graphene polarizability \cite{Hwang} $\Pi$ due to the different chiral and momentum dependence in the e-ph coupling. We have obtained an exact analytical expression for the $\mathrm{Im}\Pi^{\mathrm{ph}}$ at zero temperature \cite{myImPi}; the finite-temperature phonon self-energy is then obtained from the zero-temperature expression as \cite{Maldague}:
\begin{equation}
\mathrm{Im}\Pi^{\mathrm{ph}}(q,\omega;T,\mu_{\mathrm{c}}) = \int_0^{\infty}\mathrm{d}\mu_{\mathrm{c}}'\frac{\mathrm{Im}\Pi^{\mathrm{ph}}(q,\omega;0,\mu_{\mathrm{c}}')}{4k_{\mathrm{B}}T\mathrm{cosh}^2[(\mu_{\mathrm{c}}-\mu_{\mathrm{c}}')/2k_{\mathrm{B}}T]},
\label{eq4}
\end{equation}
here we have written out the dependence of $\mathrm{Im}\Pi^{\mathrm{ph}}$ (i.e. the imaginary part of Eq.~(\ref{eq3})) on the temperature $T$ and the chemical potential $\mu_{\mathrm{c}}(T)$ explicitly for clarity. 
$\mu_{\mathrm{c}}(T)$ is determined by requiring that the integral over all the electronic states of the Fermi function gives the electron density $n = \int\mathrm{d}\varepsilon \nu(\varepsilon)/\{\mathrm{exp}[(\varepsilon-\mu_{\mathrm{c}}(T_{\mathrm{e}}))/k_{\mathrm{B}}T_{\mathrm{e}}]+1\}$, where $\nu(\varepsilon) = g_{\mathrm{v}}g_{\mathrm{s}}\varepsilon/2\pi v^2$ is the graphene electronic density of states. Unlike in regular 2D parabolic system, a closed form for $\mu_{\mathrm{c}}$ cannot be obtained in the case for graphene, and one must numerically solve for the root of the equation which results from the integration of the Fermi function: $n = -[2(k_{\mathrm{B}}T_{\mathrm{e}})^2/\pi v^2]\mathrm{Li}_2[-\mathrm{exp}(\mu_{\mathrm{c}}/k_{\mathrm{B}}T_{\mathrm{e}})]$, where $\mathrm{Li}_2(z) = \sum_{k=1}^{\infty}(z^k/k^2)$ is the dilogarithm function. 
%

In Eq.~(\ref{eq3}), we can identify two contributions to the phonon self-energy at zero temperature $\Pi^{\mathrm{pp}} = \Pi_+^{\mathrm{pp}}+\Pi_-^{\mathrm{pp}}$; with $\Pi_+^{\mathrm{pp}}$ originating from the extrinsic carriers and therefore dependent on the Fermi level; $\Pi_-^{\mathrm{pp}}$ from the intrinsic electrons in the valence band (i.e., the `Dirac sea') and independent of the Fermi level. For optical phonons, both the extrinsic and the intrinsic parts contribute to the total power loss $P = P_{\mathrm{ext}}+P_{\mathrm{int}}$, and as the extrinsic carrier density $n$ is tuned to zero, $P_{\mathrm{ext}}$ goes to zero, but $P_{\mathrm{int}}$ remains finite even at zero extrinsic carrier density $n = 0$. The intrinsic part of the phonon self-energy $\mathrm{Im}\Pi_-^{\mathrm{pp}}$ has a simple form, which allows for an exact analytic derviation of the intrinsic power loss $P_{\mathrm{int}}$ from Eqs.~(\ref{eq2})-(\ref{eq4}):
\begin{equation}
\frac{P_{\mathrm{int}}}{N} = \frac{\omega_0^2}{12\hbar}(g_{\mathrm{op}}^{*})^2\left(\frac{\omega_0}{\varepsilon_F}\right)^2
\left[n_{B}^{\mathrm{L}}(\omega_0)-n_{B}^{\mathrm{e}}(\omega_0)\right]n_F(-\mu_{\mathrm{c}}),
\label{eq5}
\end{equation}
here $N = nA$ ($A$ is the sample area) is the number of extrinsic carriers and $(g_{\mathrm{op}}^{*})^2 = g_{\mathrm{op}}^2A/\hbar^2v^2$ the dimensionless e-ph coupling \cite{myphonon1}. Physically, $P_{\mathrm{int}}$ correponds to the power loss due to optical phonon emission through interband transitions of the valence band electrons, and is only weakly dependent on the \textit{extrinsic} carrier density through $\mu_{\mathrm{c}}$ in the Fermi function. We find that this intrinsic power loss is not small, and for $T_{\mathrm{e}} = 300\,-\,700\,\mathrm{K}$ at $n = 10^{13}\,\mathrm{cm}^{-2}$, the intrinsic power loss per unit area $P_{\mathrm{int}}/A \sim 10^3\,-\,10^5\,\mathrm{Wm}^{-2}$.  

Recent experiments \cite{ah_exp} and theory \cite{ah_th} show that $\tau_{\mathrm{ah}}$ for graphene is of the order of picoseconds, and therefore anharmonic phonon-phonon scattering occurs at a comparable (and slower) rate than e-ph scattering, causing an accumulation of non-equilibrium optical phonons (known as `hot phonons'). Some of these hot phonons are then reabsorbed back by the electron gas, thus reducing the overall power loss. We now take this effect into account by incorporating \cite{SJJ_1990} a finite phenomenological $\tau_{\mathrm{ah}}$ in our theory, and obtain the power loss as $P = -\sum_{{q}}\int{\mathrm{d}\omega}(\omega/\pi)[n_{B}^{\mathrm{L}}(\omega)-n_{B}^{\mathrm{e}}(\omega)]\mathrm{Im}\mathcal{D}(q,\omega)/[\tau(q,\omega)+\tau_{\mathrm{ah}}]$,
%
%
%
where we have written $\tau^{-1} = -2\mathrm{Im}\Pi^{\mathrm{ph}}(q,\omega)$ as the phonon damping rate due to electron-hole pair excitation.
\begin{figure}[t]
  \includegraphics[width=8.0cm,angle=0]{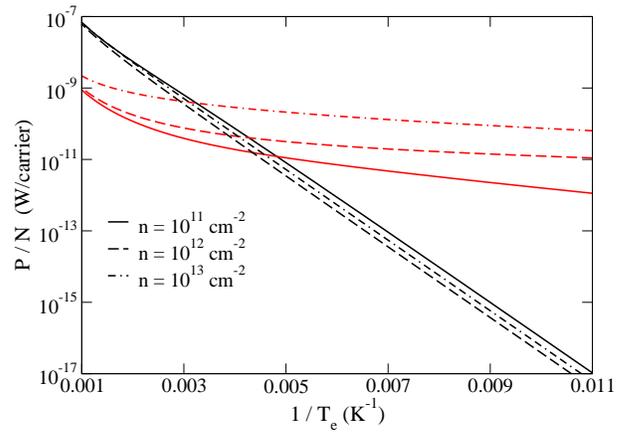}
\caption{(Color online) Power loss per carrier versus inverse electron temperature $1/T_{\mathrm{e}}$ for optical phonons (black) and acoustic phonons (red) at different electron densities $n = 10^{11}$, $10^{12}$, $10^{13}\,\mathrm{cm}^{-2}$. The slopes of the curves for optical phonons are approximately given by $\omega_0$ at low $T_{\mathrm{e}}$.} 
\label{fig1}
\end{figure}
At lower electron temperatures, electrons do not have enough energy to emit high energy optical phonons, and acoustic phonon emission becomes the dominant energy loss mechanism. The LA mode acoustic phonon in graphene at the $\Gamma$ point has an energy dispersion $\omega_q = c_{\mathrm{s}}q$ and e-ph coupling \cite{Ando} $\alpha_{{k},{k}-{q}}^{\mu\lambda} = g_{\mathrm{ac}}^2(q)[1+\mu\lambda\mathrm{cos}(\phi_{k-q}-\phi_k)]/2$, where $c_{\mathrm{s}} = 2\times10^4\,\mathrm{ms}^{-1}$ is the phonon velocity, $g_{\mathrm{ac}}(q) = Dq\sqrt{\hbar/\rho A \omega_q}$ is the coupling constant with $\rho$ being the graphene mass density, and $D = 16\,\mathrm{eV}$ the deformation potential. In contrast to the case of optical phonons, we note that $\alpha_{{k},{k}-{q}}^{\mu\lambda}$ for acoustic phonons has the same chirality and momentum dependence as in the graphene polarizability, and therefore the phonon self-energy for LA phonons is given from Eq.~(\ref{eq3}) simply by $\Pi^{\mathrm{pp}}(q,\omega) = g_{\mathrm{ac}}^2(q)\Pi(q,\omega)$, with $\Pi(q,\omega)$ being the polarizability in Ref.~\cite{Hwang}. 

As in the case for optical phonons, the acoustic phonon self-energy contains an extrinsic contribution $\Pi_+^{\mathrm{pp}}$ and an intrinsic contribution $\Pi_-^{\mathrm{pp}}$; however, we find that the intrinisc part does not contribute to the power loss and the power loss due to acoustic phonon emission originates entirely from the extrinsic contribution $P = P_{\mathrm{ext}}$. Physically, the intrinsic contribution corresponds to interband electron transitions across the conduction and the valence bands. The acoustic phonon mode, having an energy $\omega_q = c_{\mathrm{s}}q$ with $c_{\mathrm{s}}$ smaller than the graphene band velocity $v$, does not provide a possible channel for interband transition, which requires an energy greater than $vq$. Emission of acoustic phonons is therefore only possible through intraband transitions. 
%
%
%

We now calculate the power loss $P_{\mathrm{optical}}$ due to optical phonons (we use $\tau_{\mathrm{ah}} = 3.5\,\mathrm{ps}$ from Ref.~\cite{ah_th}) and $P_{\mathrm{acoustic}}$ due to acoustic phonons as a function of $T_{\mathrm{e}}$ and $n$,  
with $\mu_{\mathrm{c}}(T_{\mathrm{e}})$ determined at each $T_{\mathrm{e}}$ and the finite-temperature phonon self-energy obtained by evaluating Eq.~(\ref{eq4}). The lattice temperature is taken as zero $T_{\mathrm{L}} = 0$. Fig.~\ref{fig1} shows the power loss per carrier $P/N$ from the two contributions versus inverse electron temperature $1/T_{\mathrm{e}}$. The approximate exponential behavior of $P_{\mathrm{optical}}$ versus $1/T_{\mathrm{e}}$ is due to the fact that the electrons capable of emitting optical phonons have an amount of energy higher than $\omega_0$, which lie in the high-energy tail of the Fermi distribution with a population $\sim \mathrm{exp}(-\omega_0/k_{\mathrm{B}}T_{\mathrm{e}})$. As the phonon energy is quite high $\omega_0 = 196\,\mathrm{meV}$ in graphene, the power loss through optical phonons decrease with $T_{\mathrm{e}}$ about an order-of-magnitude faster than that in GaAs (where $\omega_0 = 36\,\mathrm{meV}$).  On the other hand, $P_{\mathrm{acoustic}}$ decreases with $T_{\mathrm{e}}$ much more slowly, lying within the range of $10^{-12}\,-\,10^{-9}\,\mathrm{W}$ for $T_{\mathrm{e}} = 100\,-\,1000\,\mathrm{K}$.

\begin{figure}[h]
  \includegraphics[width=8.0cm,angle=0]{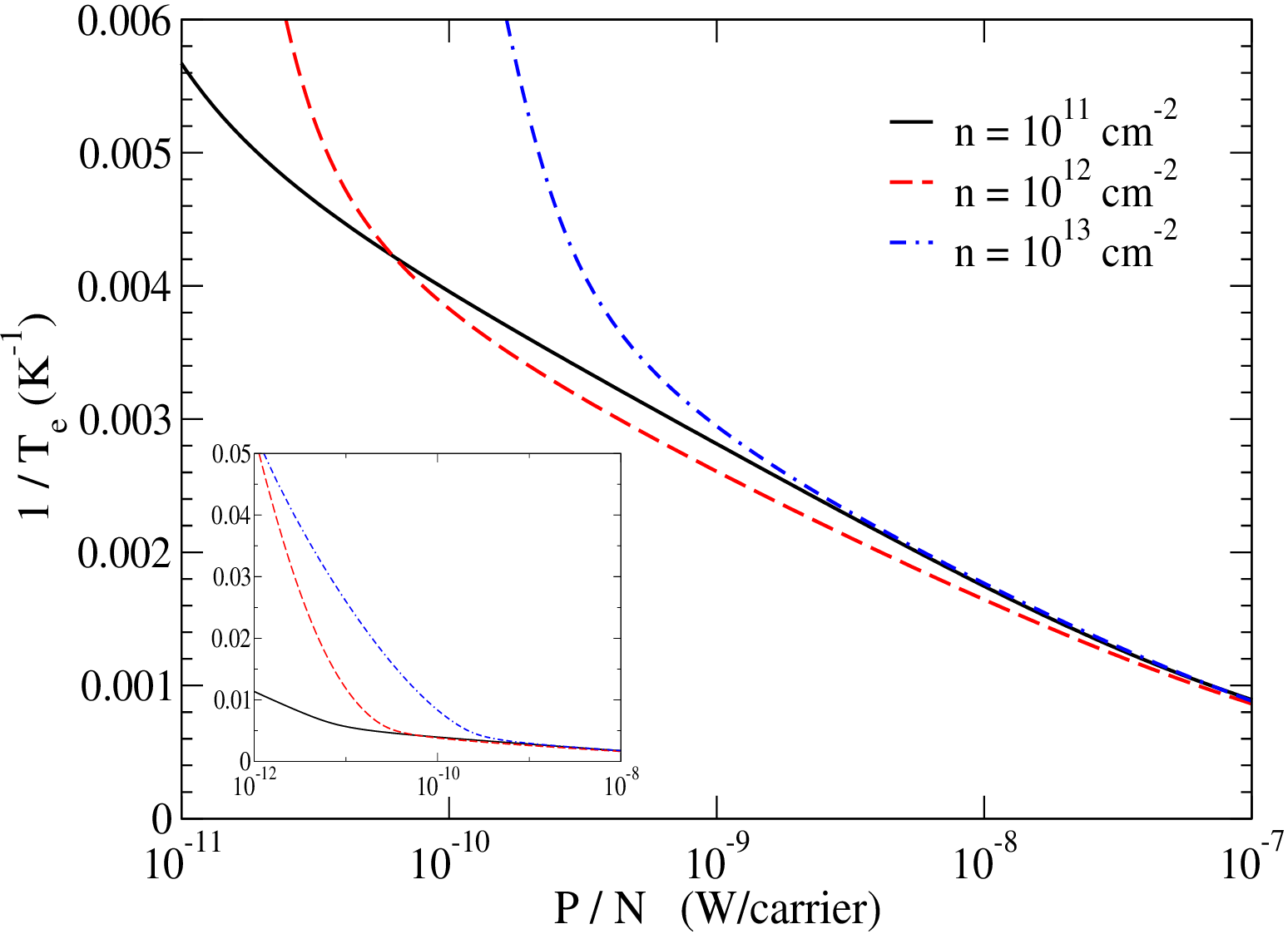}
\caption{(Color online) Inverse electron temperature $1/T_{\mathrm{e}}$ versus the total power loss per carrier $P/N$. $1/T_{\mathrm{e}}$ is plotted against $P/N$ since $P$ is the input variable in experiment (the power loss $P$ to the lattice must be equal to the experimental power input to the electrons) while $T_{\mathrm{e}}$ is the observed output. Inset: Total power loss per carrier shown within a wider temperature range down to $T_{\mathrm{e}} = 20\,\mathrm{K}$.} 
\label{fig2}
\end{figure}

%
%
The total power loss is given by the sum of the contributions from the optical phonons and acoustic phonons (Fig.~\ref{fig2}). At small values of $1/T_{\mathrm{e}}$ (high temperatures), the energy loss is predominantly through optical phonon emission, with the power loss behaving approximately exponentially. $P_{\mathrm{optical}}$ decreases as temperature is decreased, and the power loss through acoustic phonons $P_{\mathrm{acoustic}}$ becomes increasingly important. The crossover of the energy loss from predominantly optical phonon emission to predominantly acoustic phonon emission depends on the electron density, occurring at an increasing temperature with density $T_{\mathrm{e}} \sim 200\,\mathrm{K}$ for $n = 10^{11}\,\mathrm{cm}^{-2}$, $\sim 250\,\mathrm{K}$ for $10^{12}\,\mathrm{cm}^{-2}$ and $\sim 300\,\mathrm{K}$ for $10^{13}\,\mathrm{cm}^{-2}$.  

\begin{figure}[t]
  \includegraphics[width=8.0cm,angle=0]{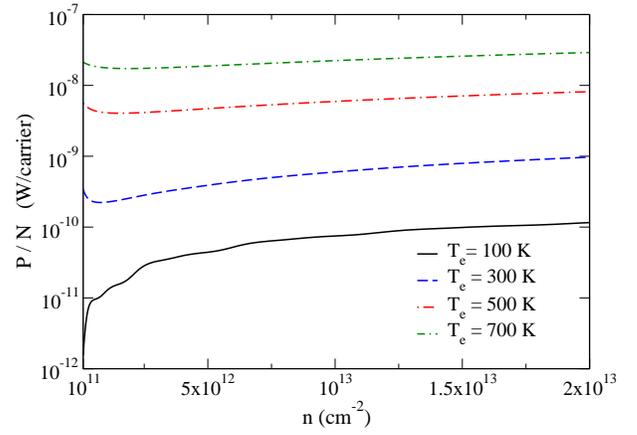}
\caption{(Color online) Total power loss per carrier $P/N$ versus electron density $n$ at different values of $T_{\mathrm{e}}$.} 
\label{fig3}
\end{figure}

We show in Fig.~\ref{fig3} the total power loss as a function of electron density $n$ for different values of $T_{\mathrm{e}}$. At $T_{\mathrm{e}} = 100\,\mathrm{K}$, the energy loss is mainly due to acoustic phonon emission, and $P/N$ increases with density. For $T_{\mathrm{e}} \gtrsim 300\,\mathrm{K}$, $P/N$ shows an upturn as $n$ is reduced towards zero, reflecting the portion of intrinsic power loss coming from optical phonon emission which must give $P/N \to \infty$ as $n\to 0$ since $P_{\mathrm{int}}$ is finite at $n = 0$. 

In conclusion, we emphasize that the intrinsic power loss has its origin from the presence of the Dirac sea in graphene, and therefore energy relaxation resulting in such intrinsic power loss occurs not just for undoped graphene with $n = 0$, but for doped graphene at all carrier densities $n > 0$ as well. We find that the total power loss per carrier taking account of both optical and acoustic phonon emission $ \sim 10^{-12}\,-\,10^{-7}\,\mathrm{W}$ for electron temperatures $T_{\mathrm{e}} \sim 20\,-\,1000\,\mathrm{K}$. Our results obtained for $T_{\mathrm{L}} = 0$ should remain valid as long as $T_{\mathrm{L}} \ll T_{\mathrm{e}}$ is satisfied in experimental situations. 

%
This work is supported by US-ONR, NSF-NRI, and SWAN SRC.


\begin{thebibliography}{18}
\bibitem{Remark1} The dephasing of the off-diagonal electron-hole coherent states described by $f_{k+-}$, $f_{k-+}$ occurs at a much faster time scale and does not contribute to the subsequent energy relaxation due to electron-phonon coupling.
\bibitem{Remark3} $g_{\mathrm{op}}$ is given by $g$ in Ref.~\cite{myphonon1}.
\end{thebibliography}
\end{document}